%% file: prl_slac.tex
\documentstyle[art12,12pt,epsf]{article}
\newcommand {\ddate}    {March 28, 1997}
\begin{document}
\title{
\begin{flushright} \normalsize
   {SLAC-PUB-7418\\
   \ddate\\\vspace{1cm}}
\end{flushright}
\begin{center} \huge \bf
Direct Measurement of\\
Leptonic Coupling
Asymmetries with Polarized Z's
\footnote{\input{sldcontract}}
\end{center}}
\author{The SLD Collaboration$^{**}$\\
Stanford Linear Accelerator Center,
Stanford University, Stanford, CA 94309}
\date{}                                                 
\maketitle
\begin{abstract}
We present direct measurements of the $Z^0$-lepton coupling
asymmetry parameters, $A_e$, $A_\mu$, and $A_\tau$, based on a data
sample of 12,063 leptonic $Z^0$ decays collected by the SLD detector.
The $Z$ bosons are produced in collisions of beams of polarized $e^-$ with
unpolarized $e^+$ at the SLAC Linear Collider.
The couplings are extracted from the measurement of the left-right and
forward-backward asymmetries for each lepton species. The results
are: 
$A_e=0.152\pm0.012\mbox{(stat)}\pm0.001\mbox{(syst)}$,
$A_\mu=0.102\pm0.034\pm0.002$, and
$A_\tau=0.195\pm0.034\pm0.003$.
\vspace*{\fill}
\begin{center} 
{\sl Submitted to Physical Review Letters}
\end{center} 
\end{abstract}
\newpage
%
\newpage

The structure of the parity violation in the electroweak interaction can
be probed directly in the production and decay of polarized $Z^0$ bosons.
All three leptonic states can be studied in $e^+e^-$ annihilations at
the $Z^0$ resonance, providing an important test of lepton universality
and the Standard Model~\cite{PDG}.

We report measurements of production and decay
asymmetries of $e^+e^-\rightarrow e^+e^-,\mu^+\mu^-, \tau^+\tau^-$
made by the SLD experiment at the SLAC Linear Collider (SLC).
SLC produces Z bosons
in $e^+e^-$ collisions using a polarized electron beam. The
polarization allows us to form the left-right cross section asymmetry
to extract the initial-state coupling $A_e$. It also enables us to
extract the final-state coupling for lepton $l$, $A_l$,
directly using the polarized forward-backward asymmetry. The
parity-violating parameter $A_l=\frac{2v_la_l}{v_l^2+a_l^2}$ depends
only on the ratio of the vector and axial vector couplings of
the lepton $l$ to the $Z^0$, $v_l$ and $a_l$ respectively.
Previous experiments at the $Z^0$ resonance \cite{LEP}
have measured the product of initial and
final-state coupling, $A_e\cdot A_l$. The polarization enables us to
present the first direct measurement of $A_\mu$. The polarized
asymmetries enhance the statistical precision on the final-state
parameter by a factor of about 25 compared to
the simple forward-backward asymmetry. The data were collected
during 1993 ($\sqrt{s}=91.26$ GeV center-of-mass energy, 63\%
$e^-$ beam polarization, $5\cdot 10^4 Z$'s analyzed) and
1994-5 ($\sqrt{s}=91.28$ GeV, 77\% polarization,
$10^5 Z$'s analyzed). The SLD detector~\cite{detector} and
the SLC accelerator~\cite{collider} have been described elsewhere.

Production of the vector Z boson in electron-positron collisions
requires the longitudinal spin projections of the electron and
positron to be parallel. The helicity of the electron controls
the direction of the spin projection of the $Z^0$. There are two
spin configurations of \linebreak the $e^+e^-$ system that have a
non-zero $Z^0$ production cross section: The cross section with
the spin of the $Z^0$ pointing in the direction opposite to the
electron's momentum is referred to as the `Left-handed' cross
section since it contains a left-handed electron. Because the
couplings of the Z boson to the fermions are parity violating,
the left-handed cross section is not equal to the right-handed
cross section, where the spin of the $Z^0$ points in the same
direction as the electron's momentum vector. In addition
to this Left-Right Asymmetry in the Z production cross section,
$A_{LR}$, Z couplings to
fermions also produce asymmetries in the angular distributions of
the Z decay products with respect to the beam axis. 


The dominant term in the cross section for polarized production
of a pair of leptons at the Z resonance is due to Z exchange:
\begin{equation}
Z(s,A_e,A_l,x)=
\left(\frac{d\sigma}{dx}\right)_{Z(s)}\hspace*{-0.2in}
=f_{Z(s)}\left((1-PA_e)(1+x^2)+(A_e-P)A_l2x\right),
\label{zterm}
\end{equation}
where $P$ is the electron
beam polarization~\cite{poldef}. $P<0$ means mostly left-handed beam
electrons and $P>0$ mostly right-handed beam electrons.
$x=\cos\ \theta$, where $\theta$ is the angle of
the $l^-$ direction with
respect to the electron beam direction. Photon exchange terms and, 
if the final-state
leptons are electrons, t-channel contributions~\cite{gterm}, have
to be taken into account. Events with
$\cos\ \theta<0$ will be called `backward' events; `forward' events
have $\cos\ \theta>0$. The sign of the polarization was switched
randomly on each beam pulse; this minimizes asymmetries
arising from SLC/SLD operation \cite{ALR}.
Half of the luminosity has a positive polarization, half a negative
polarization. We measure the polarization with the
Compton polarimeter~\cite{compton} and obtained for the 1994-5 run
$|P|=77.23\pm0.52\%\mbox{(uncertainty dominated by systematics)}$.~\cite{ALR}
The polarization for the 1993 run was $|P|=63.0\pm1.1\%$.~\cite{ALR}

Simple asymmetries can be used to extract
$A_e$ and $A_l$ from the data. The left-right asymmetry measures
the difference in $Z$ production for the left and right polarized
electron beams. The left-right forward-backward asymmetry is a
double asymmetry which is formed by taking the difference in
the number of forward ($F$) and backward ($B$) events for left
($L$) and right ($R$) beam polarization data samples. These
asymmetries can be derived from equation~(\ref{zterm}), using obvious
subscripts, (and assuming acceptance over the full solid angle) 
\begin{eqnarray}
\label{asym1}
A_{LR}&=&\frac{1}{|P|}\frac{N_L-N_R}{N_L+N_R}=A_e \\[0.1in]
\tilde{A}^l_{FB}&=&\frac{4}{3|P|}\frac{(N_{LF}-N_{LB})-(N_{RF}-N_{RB})}
{(N_{LF}+N_{LB})+(N_{RF}+N_{RB})}=A_l.
\label{asym2}
\end{eqnarray}
Events are reconstructed and accepted in this analysis only
within $|\cos\ \theta|<0.7$ where SLD trigger and reconstruction
efficiencies are high and uniform. Then the geometric factor $4/3$
in equation~(\ref{asym2}) is replaced by $1.66$.

The essence of the measurement is contained in
equations~(\ref{asym1}) and (\ref{asym2}), but instead of simply
counting events we perform a maximum likelihood fit, event by
event, using the likelihood function
\begin{equation}
{\cal L}(A_e,A_l,x)=
\int ds^\prime 
H(s,s^\prime)
\left(
Z(s^\prime,A_e,A_l,x)+Z\gamma(s^\prime,A_e,A_l,x)+\gamma(s^\prime,x)
\right)
\end{equation}
to determine simultaneously $A_e$ and $A_\mu$ with the mu-pair
events (or $A_e$ and $A_\tau$ with the tau-pair events). The
integration over $s^\prime$ is done with the program
DMIBA~\cite{dmiba} to take into account the initial-state
radiation from two times the beam energy $\sqrt{s}$ to the
invariant mass of the propagator $\sqrt{s^\prime}$
described by the radiator function $H(s,s^\prime)$. The spread
in the beam energy has a negligible effect.
The maximum likelihood fit is less sensitive to detector
acceptance as a function of polar angle than the counting method,
and has more statistical power.
$Z(...), \gamma(...)$ and Z$\gamma(...)$ are the tree-level differential
cross sections for Z exchange, photon exchange, and their
interference. The integration is performed before the fit
to obtain the coefficients $f_Z, f_{Z\gamma}$ and $f_\gamma$,
and the likelihood function becomes
\begin{equation}
{\cal L}(A_e,A_l,x)=
f_Z\cdot Z(A_e,A_l,x)+f_{Z\gamma}\cdot Z\gamma(A_e,A_l,x)+
f_\gamma\cdot\gamma(x).
\end{equation}
These coefficients give the relative sizes of the three terms at
the SLC center-of-mass energy. For the electron final-state we include the
t-channel contributions in the likelihood function to determine
$A_e$.


Leptonic decays of
the $Z$ are characterized by low charged multiplicity and two
back-to-back leptons (or in the case of the
tau-pair events, the tau decay products). This analysis relies
on the charged track reconstruction in the central drift
chamber (CDC) and the measurement of the energies associated
with the tracks in the liquid argon calorimeter (LAC). A
pre-selection requires lepton-pair events to have between 2
and 8 charged tracks, each of which must pass
within 1 cm of the nominal $e^+e^-$ interaction point. This
excludes most hadronic $Z$ decays, which have an average
charged multiplicity of approximately 20. Since the leptons
have about 45 GeV in energy, there is little problem assigning
reconstructed tracks to one of two event hemispheres,
corresponding to the two leptons. One hemisphere must have a net
charge 1 and the other a net charge -1 to ensure unambiguous
assignment of the scattering
angle. Each event is assigned a polar production angle based on
the thrust axis defined by the charged tracks. Additional
requirements are imposed to select $e^+e^-$, $\mu^+\mu^-$, and
$\tau^+\tau^-$ final-states and further reduce backgrounds.
Table~\ref{sample} summarizes the electron,
muon, and tau event selections.

A single additional cut is required to select the $e^+e^-$ final state.
We require the sum of the energy deposited in the LAC
by the highest momentum track in each hemisphere to be more than
45 GeV. The electron sample has a small contamination (0.7\%)
from tau events.


Events of the type $Z\rightarrow\mu^+\mu^-$ must have
the invariant mass of the measured charged tracks above
70 GeV/c$^2$. This removes most $Z\rightarrow\tau^+\tau^-$ events and
virtually all two-photon events and any remaining hadronic Z decays.
The majority of events remaining are $Z\rightarrow\mu^+\mu^-$ and
$e^+e^-\rightarrow e^+e^-$. We remove the $e^+e^-$ final-state by
requiring the energy deposited in the LAC by the highest momentum
track in each hemisphere to be less than 10 GeV (but more than 0 GeV to
eliminate events with both tracks entering spaces between calorimeter
modules). The muon sample has no backgrounds except for 0.4\% tau
events.


The tau selection takes the complement of the muon sample and requires
the event mass to be less than 70 GeV/c$^2$. Requiring the maximum of the
two energies in the LAC associated\linebreak
to the highest momentum track in each
hemisphere to be non-zero but below 27.5 GeV removes
$e^+e^-\rightarrow e^+e^-$ events. Two-photon
events are suppressed by requiring the angle between the momenta of the
two hemispheres, as measured by the sum of the charged track momenta in
each hemisphere, to be greater than $160^0$.  Requiring one charged
track to have momentum greater than 3 GeV/c also reduces two-photon
background. The remaining background from hadronic Z decays is suppressed
by requiring each hemisphere invariant mass, measured using charged
tracks, to be less than 1.8 GeV/c$^2$. The tau sample is contaminated
with muons (2\%), electrons (1.5\%), two-photon (1\%), and hadronic
events (0.5\%).


There are several systematic effects which can bias the result. (1)
The uncertainty on the beam polarization is correlated among all the
measurements and corresponds to an uncertainty on $A_e$ and
$A_l$ of $\pm 0.001$.
(2) Uncertainties in the amount of background and its effect on the
fitted parameters must also be taken into account.
For the $e^+e^-$ and $\mu^+\mu^-$ final-states we rely on Monte Carlo
simulation to estimate the effect of backgrounds. For these
samples, background has a negligible effect ($<0.0005$).
The tau sample contains significant background, which we have studied
using samples of background-rich events selected from the data itself.
These events were used to estimate the polar-angle distribution
of the background events that eventually populate the tau sample.
The results are listed in table~\ref{systematics}. (3) The 
dominant systematic error in the tau analysis comes about
because we measure not the taus themselves, but their decay
products. The helicities of the two taus from Z decay are 100\%
anti-correlated: one will be left-handed and the other
right-handed. So, given the V-A structure of tau decay~\cite{tsai},
the decay products from the $\tau^+$ and the $\tau^-$ from a
particular $Z$ decay will take their energies from the same set
of spectra.  For example, if both taus decay to $\pi\nu$, then
both pions will generally be low in energy (in the case of a
left-handed $\tau^-$ and right-handed $\tau^+$) or both will be
generally higher in energy. The effect is strong at SLD because
the high beam polarization induces very high and asymmetric tau
polarization as a function of polar production angle. In addition,
the sign of the polarization is approximately opposite for left- and
right-handed  $e^-$ beam events at a given polar angle.
Thus selecting events based
on event mass, for example, may cause polar angle dependence in
selection efficiency for taus which has opposite effect for
taus from events produced with the left and right polarized
electron beam. Taking all tau decay modes into account, using
Monte Carlo simulation, we find an overall shift of
$+0.0080\pm0.0019$ on $A_\tau$ due to this effect (the value
extracted from the fit must be reduced by this amount). The
value of $A_e$ extracted from $\tau^+\tau^-$ final-states is
not affected since the overall relative efficiencies for
left-beam and right-beam events are not changed significantly
(only the polar angle dependence of the efficiencies are
changed). (4) The calculation of the maximum likelihood function
depends on the average beam energy $\sqrt{s}$. The uncertainty
due to a $\pm1\sigma$ variation of this energy is of the order
$10^{-3}$ (see table~\ref{systematics}).

We have also studied the effect of the uncertainty in the
thrust axis determination and found that this contribution is
negligible. The selection efficiency as a function of polar
angle is another possible source of bias in $A_l$.
If this efficiency is symmetric about $\cos\ \theta=0$ then
$A_l$ (and $A_e$) will be unaffected for muons and taus (see
equations~(\ref{asym1}) and (\ref{asym2})). However, the maximum
likelihood fit for the $e^+e^-$ final state will be affected even for
a symmetric efficiency, if it is not uniform. We did not find a
significant deviation from a uniform efficiency
within $|\cos\ \theta|<0.7$ and
estimate conservatively the upper limit of $\Delta A_e<10^{-4}$.
A small detector-induced forward-backward asymmetry
would also introduce a bias in $A_l$,
but $A_e$ would still be unaffected. Using the data,
we have studied the effect of the selection cuts as a function
of polar angle. No systematic effect is observed and we assign
a conservative systematic uncertainty of $\Delta A_l=5\cdot10^{-4}$.
The systematic uncertainties are summarized in
table~\ref{systematics}, they are negligible compared to the
statistical uncertainties.

\begin{figure}[p]
\centering
\epsfxsize=3.1in
\centerline{\epsffile{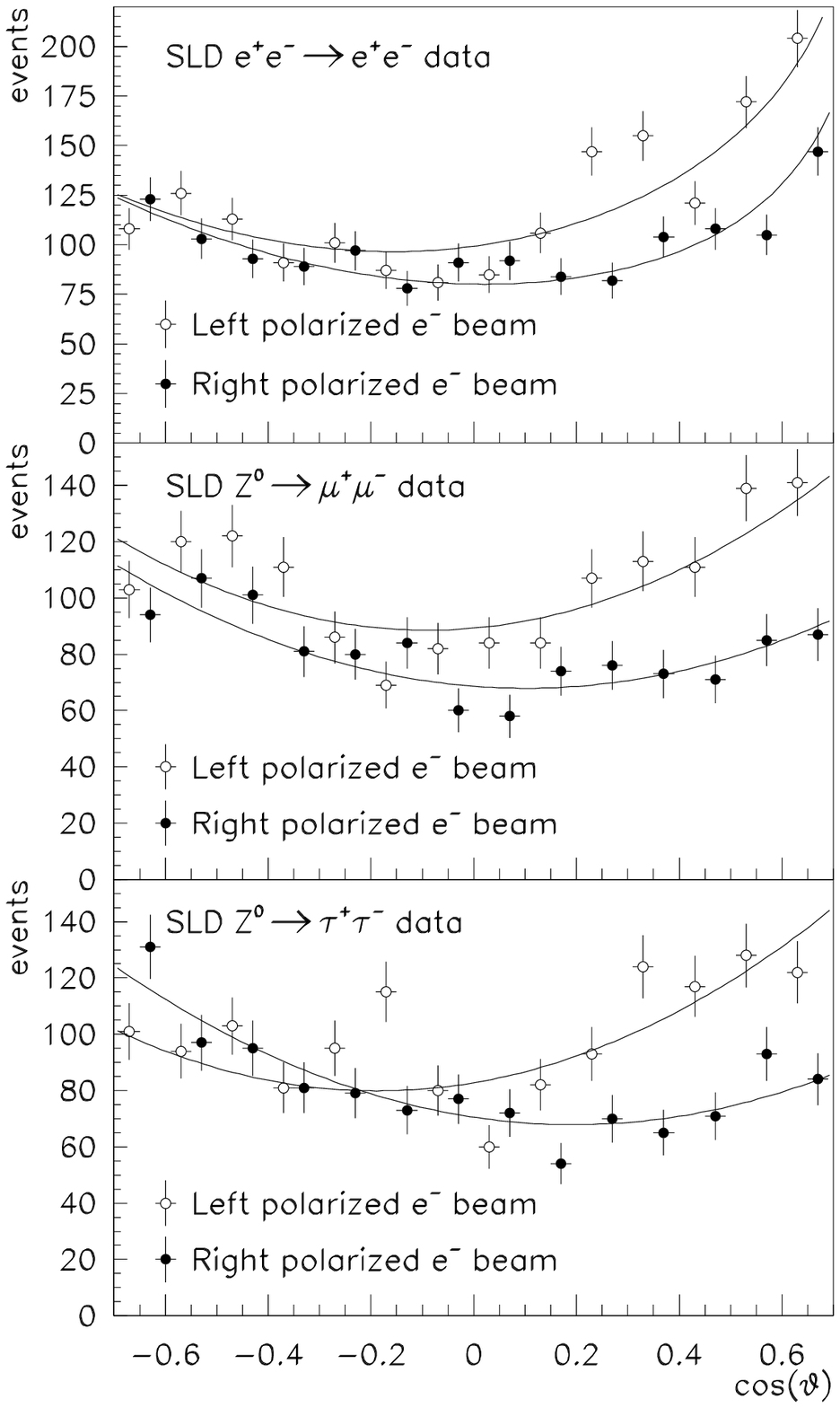}}
\caption{Polar angle distribution for $Z$ decays to $e$, $\mu$ and $\tau$
pairs for the 1994-5 SLD run. The asymmetries in the 1993 data look similar
but are less pronounced due to the lower polarization.}
\label{distr}
\end{figure}

Figure~(\ref{distr}) shows the $\cos\ \theta$ distributions for electron,
muon and tau final-states for the 1994-5 data.  The solid line represents
the fit, while the points with error bars show the data in bins of 0.1
in $\cos\ \theta$.

We have presented direct measurements of the Z-lepton coupling
asymmetries $A_e$, $A_\mu$, and $A_\tau$ using
$e^+e^-\rightarrow e^+e^-, \mu^+\mu^-, \tau^+\tau^-$ events
produced with a polarized $e^-$ beam. The results are
\begin{equation}
\matrix{
A_e=0.152\pm0.012     \cr
A_\mu=0.102\pm0.034   \cr
A_\tau=0.195\pm 0.034. }
\end{equation}
Our results are consistent with lepton universality. Assuming
universality, we can combine them into $A_{e-\mu-\tau}$ which
in the context of the standard model is simply related to the
electroweak mixing angle~\cite{PDG}
\[
\matrix{
A_{e-\mu-\tau}=0.151\pm0.011,}
\]
\begin{equation}
\sin^2\theta^{\mbox{\it eff, lept}}_W\equiv
\frac{1}{4}\left(1-\frac{v_l}{a_l}\right)=0.2310\pm0.0014.
\end{equation}
This measurement is independent of the SLD result from
$A_{LR}$~\cite{ALR} using Z decays to hadrons. The combined
results from the four LEP experiments~\cite{LEP} can be written
as $A_e=0.1461\pm0.0059$, $A_\mu=0.1476\pm0.0132$,
$A_\tau=0.1463\pm0.0062$.


We thank the personnel of the SLAC accelerator department and the
technical staffs of our collaborating institutions for their outstanding
efforts on our behalf. This work was supported by the Department of
Energy, the National Science Foundation, the Istituto Nazionale di
Fisica Nucleare of Italy, the Japan-US Cooperative Research Project
on High Energy Physics, and the Science and Engineering Research
Council of the United Kingdom.


%
%

\begin{table}[hbtp]
\caption{Summary of event selections for $Z\rightarrow l^+l^-$}
\label{sample}
\begin{center}
\begin{tabular}{|llcl|}
\hline
event  & backgr. as \%  & effic. in            & \# of selected \\
sample & of sel. events & $|\cos\ \theta|<0.7$ & events\\ \hline
$e^+e^-\rightarrow e^+e^-$ &
   0.7\% $\tau^+\tau^-$ &      & 1993 run: \hspace{0.05in} 1434 \\
 &                      & 92\% & 1994-5 run: 3093 \cr \hline
$Z\rightarrow\mu^+\mu^-$   &
   0.4\% $\tau^+\tau^-$ &      & 1993 run: \hspace{0.05in} 1185 \\
 &                      & 96\% & 1994-5 run: 2603 \\ \hline
$Z\rightarrow\tau^+\tau^-$ & 
   1.5\% $e^+e^-$       &      & \cr
 & 2\% $\mu^+\mu^-$    &      & 1993 run: \hspace{0.05in} 1211 \\
 & 1\% two-photon       & 90\% & 1994-5 run: 2537 \\
 & 0.5\%  hadrons              &      & \\
\hline
\end{tabular}
\end{center}
\end{table}

\begin{table}[p]
\caption{Uncertainties on $A_e$, $A_\mu$ and $A_\tau$ in the polarized
asymmetry analysis of leptonic Z decays. The total systematic error
is the quadratic sum of the systematic contributions.}
\label{systematics}
\begin{center}
\begin{tabular}{|lccc||cccc|}
\hline
         & total & stat. & total & \multicolumn{4}{l|}{Systematic contributions}
\\ \cline{5-8}
         &       &       & syst. & $e^-$ pol. & backgr.&effic. bias&$\sqrt{s}$\\
\hline
$A_e$    & 0.012 & 0.012 & 0.001 & 0.001      & negl.   & negl.       & 0.001\\
$A_\mu$  & 0.034 & 0.034 & 0.002 & 0.001      & negl.   & n.a.        & 0.002\\
$A_\tau$ & 0.034 & 0.034 & 0.003 & 0.001      & 0.001   & 0.002       & 0.001\\
\hline
\end{tabular}
\end{center}
\end{table}

%
%

\clearpage

\section*{**List of Authors} 
\input{sldauthor} 

\end{document}

%% file: sldcontract.tex
This work was supported by Department of Energy
  contracts:
  \hbox{DE-FG02-91ER40676 (BU),}
  DE-FG03-92ER40701 (CIT),
  DE-FG03-91ER40618 (UCSB),
  DE-FG03-92ER40689 (UCSC),
  DE-FG03-93ER40788 (CSU),
  DE-FG02-91ER40672 (Colorado),
  DE-FG02-91ER40677 (Illinois),
  DE-AC03-76SF00098 (LBL),
  DE-FG02-92ER40715 (Massachusetts),
  DE-AC02-76ER03069 (MIT),
  DE-FG06-85ER40224 (Oregon),
  DE-AC03-76SF00515 (SLAC),
  DE-FG05-91ER40627 (Tennessee),
  DE-AC02-76ER00881 (Wisconsin),
  DE-FG02-92ER40704 (Yale);
  National Science Foundation grants:
  PHY-91-13428 (UCSC),
  PHY-89-21320 (Columbia),
  PHY-92-04239 (Cincinnati),
  PHY-88-17930 (Rutgers),
  PHY-88-19316 (Vanderbilt),
  PHY-92-03212 (Washington);
  the UK Science and Engineering Research Council
  (Brunel and RAL);
  the Istituto Nazionale di Fisica Nucleare of Italy
  (Bologna, Ferrara, Frascati, Pisa, Padova, Perugia);
  and the Japan-US Cooperative Research Project on High Energy Physics
  (Nagoya, Tohoku).

%% file: sldauthor.tex
%
%
%
  \def\iADEL{$^{(1)}$}
  \def\iBOL{$^{(2)}$}
  \def\iBU{$^{(3)}$}
  \def\iBRUN{$^{(4)}$}
  \def\iUCSB{$^{(5)}$}
  \def\iUCSC{$^{(6)}$}
  \def\iCIN{$^{(7)}$}
  \def\iCSU{$^{(8)}$}
  \def\iCOLO{$^{(9)}$}
  \def\iCOL{$^{(10)}$}
  \def\iFER{$^{(11)}$}
  \def\iFRA{$^{(12)}$}
  \def\iILL{$^{(13)}$}
  \def\iLBL{$^{(14)}$}
  \def\iMIT{$^{(15)}$}
  \def\iMASS{$^{(16)}$}
  \def\iMISS{$^{(17)}$}
  \def\iMOSC{$^{(18)}$}
  \def\iNAG{$^{(19)}$}
  \def\iOREG{$^{(20)}$}
  \def\iPAD{$^{(21)}$}
  \def\iPERU{$^{(22)}$}
  \def\iPISA{$^{(23)}$}
  \def\iRUT{$^{(24)}$}
  \def\iRAL{$^{(25)}$}
  \def\iSOGANG{$^{(26)}$}
  \def\iSOONG{$^{(27)}$}
  \def\iSLAC{$^{(28)}$}
  \def\iTENN{$^{(29)}$}
  \def\iTOH{$^{(30)}$}
  \def\iVAND{$^{(31)}$}
  \def\iWASH{$^{(32)}$}
  \def\iWISC{$^{(33)}$}
  \def\iYALE{$^{(34)}$}
  \def\dead{$^{\dag}$}
  \def\andgen{$^{(a)}$}
  \def\andper{$^{(b)}$}
%
%
\begin{center}
\mbox{K. Abe                 \unskip,\iNAG}
\mbox{K. Abe                 \unskip,\iTOH}
\mbox{T. Akagi               \unskip,\iSLAC}
\mbox{N.J. Allen             \unskip,\iBRUN}
\mbox{W.W. Ash               \unskip,\iSLAC$^\dagger$}
\mbox{D. Aston               \unskip,\iSLAC}
\mbox{K.G. Baird             \unskip,\iRUT}
\mbox{C. Baltay              \unskip,\iYALE}
\mbox{H.R. Band              \unskip,\iWISC}
\mbox{M.B. Barakat           \unskip,\iYALE
% ...
}
\it
  \vskip \baselineskip                   
  \vskip \baselineskip                   
%
%
%
  \iADEL
     Adelphi University,
     Garden City, New York 11530 \break
  \iBOL
     INFN Sezione di Bologna,
     I-40126 Bologna, Italy \break
  \iBU
     Boston University,
     Boston, Massachusetts 02215 \break
  \iBRUN
     Brunel University,
     Uxbridge, Middlesex UB8 3PH, United Kingdom \break
  \iUCSB
     University of California at Santa Barbara,
     Santa Barbara, California 93106 \break
  \iUCSC
     University of California at Santa Cruz,
     Santa Cruz, California 95064 \break
  \iCIN
     University of Cincinnati,
     Cincinnati, Ohio 45221 \break
  \iCSU
     Colorado State University,
     Fort Collins, Colorado 80523 \break
  \iCOLO
     University of Colorado,
     Boulder, Colorado 80309 \break
  \iCOL
     Columbia University,
     New York, New York 10027 \break
  \iFER
     INFN Sezione di Ferrara and Universit\`a di Ferrara,
     I-44100 Ferrara, Italy \break
  \iFRA
     INFN  Lab. Nazionali di Frascati,
     I-00044 Frascati, Italy \break
  \iILL
     University of Illinois,
     Urbana, Illinois 61801 \break
  \iLBL
     E.O. Lawrence Berkeley Laboratory, University of California,
     Berkeley, California 94720 \break
  \iMIT
     Massachusetts Institute of Technology,
     Cambridge, Massachusetts 02139 \break
  \iMASS
     University of Massachusetts,
     Amherst, Massachusetts 01003 \break
  \iMISS
     University of Mississippi,
     University, Mississippi  38677 \break
  \iMOSC
    Moscow State University,
    Institute of Nuclear Physics
    119899 Moscow, Russia    \break
  \iNAG
     Nagoya University,
     Chikusa-ku, Nagoya 464 Japan  \break
  \iOREG
     University of Oregon,
     Eugene, Oregon 97403 \break
  \iPAD
     INFN Sezione di Padova and Universit\`a di Padova,
     I-35100 Padova, Italy \break
  \iPERU
     INFN Sezione di Perugia and Universit\`a di Perugia,
     I-06100 Perugia, Italy \break
  \iPISA
     INFN Sezione di Pisa and Universit\`a di Pisa,
     I-56100 Pisa, Italy \break
  \iRUT
     Rutgers University,
     Piscataway, New Jersey 08855 \break
  \iRAL
     Rutherford Appleton Laboratory,
     Chilton, Didcot, Oxon OX11 0QX United Kingdom \break
  \iSOGANG
     Sogang University,
     Seoul, Korea \break
  \iSOONG
     Soongsil University,
     Seoul, Korea  156-743 \break
  \iSLAC
     Stanford Linear Accelerator Center, Stanford University,
     Stanford, California 94309 \break
  \iTENN
     University of Tennessee,
     Knoxville, Tennessee 37996 \break
  \iTOH
     Tohoku University,
     Sendai 980 Japan \break
  \iVAND
     Vanderbilt University,
     Nashville, Tennessee 37235 \break
  \iWASH
     University of Washington,
     Seattle, Washington 98195 \break
  \iWISC
     University of Wisconsin,
     Madison, Wisconsin 53706 \break
  \iYALE
     Yale University,
     New Haven, Connecticut 06511 \break
  \dead
     Deceased \break
  \andgen
     Also at the Universit\`a di Genova \break
  \andper
     Also at the Universit\`a di Perugia \break
\rm
%
\end{center}
\centerline{(SLD Collaboration)}

%% file: prl_slac.bbl
\begin{thebibliography}{9}
\bibitem{poldef}We define the polarization as $P=\frac{R-L}{R+L}$ where
$L$ ($R$) is the number of left-handed (right-handed) electrons.
\bibitem{gterm} The photon s-channel cross section terms are
$\left(\frac{d\sigma}{dx}\right)_{\gamma(s)}\hspace*{-0.2in}\propto(1+x^2)$
and
$\left(\frac{d\sigma}{dx}\right)_{\gamma(s)Z(s)}\hspace*{-0.2in}\propto
\frac{v_e}{a_e}\frac{v_l}{a_l}\times
\left(\left(1-P\frac{a_e}{v_e}\right)\left(1+x^2\right)+
\left(\frac{a_e}{v_e}-P\right)\frac{a_l}{v_l}2x\right)$,
the dominant t-channel cross section for electrons is
$\left(\frac{d\sigma}{dx}\right)_{\gamma(t)}\hspace*{-0.2in}\propto
\frac{4+(x+1)^2}{(x-1)^2}$
\end{thebibliography}
